# Nonlinear Participation Factor-based Power System Model Reduction Addressing Near-Resonance Conditions

Mahsa Sajjadi, *Student member, IEEE*, Kai Sun, *Fellow, IEEE*

***Abstract*— This paper proposes an adaptive model reduction approach based on nonlinear participation factors (NPFs) which determines the most effective selection of unimportant generators in a power system to be linearized to accelerate time-domain simulation for the entire system. The method enables a dynamic transition between the full-order model, and a hybrid reduced model. It uses modal energies to rank system modes under contingencies, and computes NPFs for highly energized modes based on Normal Form theory. To accelerate the computation of NPFs and make it achievable online, a tensor contraction technique is introduced. The proposed approach is tested on a 48-machine, 140-bus NPCC system using both partitioned and unpartitioned strategies. It demonstrates significant simulation speedup while preserving better accuracy than a linear participation factor-based method if the nonlinear behaviors of the system cannot be ignored, especially when a near-resonance condition is presented.**

***Index Terms*-- Model reduction, nonlinear participation factors, normal form, power system simulation.**

## I. Introduction

POWER system simulation has been crucial for many years in evaluating stability and dynamic performance following disturbances, ensuring reliable operation and resilience of the grid. A challenging task for power system planning and operation studies is to keep the operating point within a reliable and safe range to prevent system instabilities. Faster than real time awareness of possible system instability is important because it enables system operators to take prompt corrective action to maintain a reliable and stable operating condition. Power system model reduction is a widely used tool to accelerate online simulation for a power system. The aim of model reduction is to simulate faster and accurate dynamic models that can preserve the most important characteristics of the real power system [1]-[3].

There are two types of model reduction techniques: linear model reduction and nonlinear model reduction techniques [1], [2]. Nonlinear modal interactions occur more frequently than ever due to growing complexity and uncertainty in the dynamics of modern power grids, especially in high-stress operating circumstances or when the system is subjected to a large disturbance [4]-[6]. A power grid in a near-resonance condition is one example where this might happen [7]. In such circumstances, linear model reduction techniques might not give an appropriate representation of the power system model because power system non-linear behavior cannot be fully captured by a reduced model obtained by linear model reduction. This makes methodologies based on nonlinear model reduction an appealing idea for improving the comprehension of power system dynamics.

Many model reduction methods have been proposed in the literature and applied to the power system analysis, including singular perturbation [8], [9], modal reduction [10], [11], balanced truncation [12]-[15], coherency [16], [17], and reduction by moment matching [18], [19]. While methods such as balanced truncation and moment matching are effective in reducing system order in a transformed space, approaches based on singular perturbation, modal analysis and participation factor (PF) analysis provide a different perspective by focusing on the level of interaction between state variables and system modes. PFs are dimensionless indices, and are widely used in many different fields, including electrical power system [20], [21]. To speed up power system simulations, for example, assessing and ranking the PFs for each component in a power grid can assist minimize the grid model and concentrate more on the highly participating components [20]. One advantage of using participation factors is that they assess mode-machine relationships without being affected by eigenvector scaling. This is because participation factors are derived from both the left and right eigenvectors[20], [22].

Papers [23]-[28] performed model reduction by singular perturbation based on a separation between fast and slow modes using modal analysis and linear PFs to accelerate the power system simulation. The methods in [23], [24] use linear PFs to form disjoint groups of eigenvalues and their associated state variables. By grouping these modes and states, a reduced model is created by removing specific state groups. Reduced converter models for single converter units were first developed in [23] and later expanded in [24] to include converter interconnections. This was achieved by

M. Sajjadi, and K. Sun are with the Department of EECS, University of Tennessee, Knoxville, TN (Emails: msajjad1@vols.utk.edu, and kaisun@utk.edu).

sequentially eliminating the fastest modes using linear participation factors. This approach does not directly assess the importance of each state individually. Using a similar methodology, [25] introduced an approach based on PFs calculation and the analysis of interaction modes between power electronics devices and a particular line to determine the appropriate level of model complexity required for the transmission lines in the system from the stability perspective.

Ref. [26] demonstrates that in systems with insufficient time-scale separation, certain fast dynamics with small participation factors, yet interacting with slow dynamics, cannot be disregarded. The participation of these fast states must be retained in the slow sub-model. The study in [26] does not explicitly illustrate the characteristics of these coupling dynamics. Similar approaches are presented in [27], [28] for model reduction of multi-timescale systems using singular perturbation and linear PFs.

In our previous work in [20], model reduction was performed using linear PFs analysis, without accounting for the high nonlinearity of the system. The results were then compared with two other approaches: one involving a fully linearized model reduction and another using model reduction based on the rotor angle deviation criterion. The results showed that the reduced model obtained through linear PFs analysis exhibited lower errors compared to the two other methods.

To address the high nonlinearity or near resonance conditions, this paper proposes a model reduction framework utilizing nonlinear participation factors (NPFs) of second orders or higher obtained from Normal Form theory to identify the most effective selection of generating units for linearization, thereby accelerating time-domain simulations. Normal form is used for understanding resonance phenomena and modal interactions, which are critical for stability analysis [4]-[6], [29]. PFs calculation does not require prior knowledge of a specific contingency, and if a significant change in system conditions occurs (such as a major topology change due to a permanent fault), the analysis can be updated accordingly. The tensor contraction technique significantly accelerates NPF computation in this study, making online implementation of the proposed approach feasible if real-time updates are needed (e.g., during major system changes). While PFs are derived from pre-fault conditions, their utility extends beyond specific contingencies, providing a generalized framework for assessing modal interactions and system behavior. For permanent topology changes, the operating point can be updated without impacting computational efficiency over extended periods.

In the proposed approach, each mode's energy is determined, and the system's modes are arranged in order of increasing energy. For highly energetic modes, the second order nonlinear participation factors are then calculated. The simulation results are presented for both partitioned and unpartitioned cases.

The main contributions of this paper are as follows. First, a novel adaptive model reduction approach is proposed to address near-resonance conditions or when nonlinearities in the system's response cannot be ignored. This approach allows dynamic switching between the detailed model and reduced model, including a fully nonlinear model and a hybrid model. The hybrid model selects unimportant generator functions to linearize while retaining the nonlinear details of the generators that significantly participate in the mode of interest. Second, this approach ranks the modes of the system under contingencies by modal energies and calculates the linear and nonlinear PFs for highly energized modes. Third, using a new tensor contraction algorithm, the computation of Normal Form terms and NPFs is significantly accelerated. This avoids costly matrix operations and nested loops, making the method ideal for large-scale power system stability analysis where nonlinear effects play a crucial role. Fourth, the model reduction results are presented for both partitioned and unpartitioned system. The results obtained from linear PFs and nonlinear PFs are also compared.

In the rest of this paper, Section II briefly describes Normal Form theory and second order nonlinear participation factor calculation. In Section III, the model reduction process is explained. Section IV includes the simulation results of the proposed approach on NPCC system for partitioned and unpartitioned cases. Finally, the conclusion is drawn in Section V.

## II. Normal Form and Nonlinear Participation factor

This section presents an overview of Normal Form theory and nonlinear PFs.

### A. Normal Form

The fundamental idea of Normal Form theory is to identify an analytical change in coordinates using the origin as a fixed point, making it easier to investigate the vector field in terms of the new variables [4], [6].

Consider a general first-order ordinary differential equation for nonlinear dynamical system described as follows:

$$\dot{x} = f(x) \tag{1}$$

where $x$ is the vector of system states, $u$ is the inputs and $f(x)$ is a nonlinear function. At the system equilibrium point, all state variables are constant, and their derivatives should be zero.

By computing the Jacobian matrix $A$ of (1), which represents the first-order derivatives of the system's state variables with respect to time, the linear approximation for the dynamics of the system can be derived. The system's equations around the equilibrium point are expanded in a second-order Taylor series to provide more detailed understanding of the local behavior around the equilibrium point as follows:

$$\dot{x}_i = A_i x + \frac{1}{2} X^T H^i X + H.O.T. \tag{2}$$

where $A_i$ is the $i^{th}$ row of Jacobian matrix $A$ and $H^i$ is the Hessian matrix with second-order derivatives. By a linear transformation $x = Uy$, (2) will be transformed to the Jordan form as:

$$\dot{y} = \Lambda y + \frac{1}{2} V \begin{bmatrix} y^T U^T H^1 U\, y \\ y^T U^T H^2 U\, y \\ \dots \\ y^T U^T H^n U\, y \end{bmatrix} \tag{3}$$

where $y$ is the vector of Jordan form coordinates, $U$ and $V$ are right and left eigenvectors, respectively [4], [29]. $\Lambda$ is a diagonal matrix, consisting of eigenvalues. The state equation for the $j^{th}$ Jordan form variable is expressed as follows:

$$\dot{y}_j = \lambda_j y_j + \sum_{k=1}^{n} \sum_{l=1}^{n} C_{kl}^{j}\, y_k\, y_l \tag{4}$$

where $C^j_{kl}$ is the $k^{\text{th}}$ row and $l^{\text{th}}$ column of matrix $C^j$ and it calculated as follows:

$$C^j = \frac{1}{2} \sum_{p=1}^{n} V_{jp} U^T H^p U \tag{5}$$

The Normal Form of (4) is obtained through a nonlinear coordinate transformation:.

$$y_j = z_j + \sum_{k=1}^{n} \sum_{l=1}^{n} h_{2\,jkl}\, z_k\, z_l \tag{6}$$

where $z$ is the Normal Form coordinate system and $h_2$ is the second-order transformation terms as follows:

$$h_{2\,jkl} = C_{kl}^{j} \big/ \left( \lambda_k + \lambda_l - \lambda_j \right) \quad \lambda_k + \lambda_l - \lambda_j \neq 0 \tag{7}$$

According to Normal Form theory, if no second-order resonances are present, a system that includes both first and second-order terms can be reduced to a purely linear system in its Normal Form. The Normal Form of system in $z$-coordinate is:

$$\dot{z} = \Lambda z + O\left( \|z\|^3 \right) \tag{8}$$

The system in (8) can be described in the uncoupled form if the higher order terms are neglected:

$$\begin{cases} \dot{z}_j = \lambda_j z_j \\ z_j(0) = z_{jo} \end{cases} \tag{9}$$

The closed form solutions for the system in the Jordan and physical spaces using equations (1)-(9) become:

$$y_j(t) = z_{jo} e^{\lambda_j t} + \sum_{k=1}^{n} \sum_{l=k}^{n} h_{2\,jkl}\, z_{ko}\, z_{lo}\, e^{(\lambda_k + \lambda_l)t} \tag{10}$$

$$x_i(t) = \sum_{j=1}^{n} u_{ij}\, z_{jo} e^{\lambda_j t} + \sum_{j=1}^{n} u_{ij} \left[ \sum_{k=1}^{n} \sum_{l=k}^{n} h_{2\,jkl}\, z_{ko}\, z_{lo}\, e^{(\lambda_k + \lambda_l)t} \right] \tag{11}$$

where $z_{jo}$ is the initial condition [4]-[6].

### B. Nonlinear Participation Factor

Participation factors indicate the magnitude of modal oscillations in a machine state when only that specific machine state is perturbed. In this sense, the initial condition vector $x_0 = e_k = [0\ 0\ \dots 1|_k \dots\ 0]$ is applied, and the initial conditions in Jordan form and in the $z$-coordinate become:

$$y_{jo} = v_{ji} \tag{12}$$

$$z_{jo} = y_{jo} - \sum_{p=1}^{n} \sum_{q=p}^{n} h_{2\,jpq} v_{pi}\, v_{qi} = v_{ji} + v_{2\,jii} \tag{13}$$

By inserting (13) to (11) and simplifying, the expressions for the NPFs (nonlinear PFs) can be written as:

$$x_i(t) = \sum_{j=1}^{n} p_{2ij} e^{\lambda_j t} + \sum_{k=1}^{n} \sum_{l=k}^{n} p_{2ikl} e^{(\lambda_k + \lambda_l)t} \tag{14}$$

$$p_{2ij} = u_{ij}(v_{ji} + v_{2\,jii}) = u_{ij} v_{ji} + u_{ij} v_{2\,jii} \overset{\text{def}}{=} p_{2ij_L} + p_{2ij_{NL}} \tag{15}$$

$$p_{2ikl} = u_{2ikl}(v_{ki} + v_{2kii})(v_{li} + v_{2lii}) \tag{16}$$

And

$$v_{2\,jii} = -\sum_{p=1}^{n} \sum_{q=p}^{n} h_{2\,jpq} v_{pi}\, v_{qi} \tag{17}$$

$$u_{2ikl} = \sum_{j=1}^{n} h_{2\,jkl}\, u_{ij} \tag{18}$$

where $p_{2ij}$ is the second-order participation of the $j^{th}$ single-eigenvalue mode in the $i^{th}$ state and $p_{2ikl}$ is the second-order participation of the $i^{th}$ state in the mode formed by the combination of the $k^{th}$ and $l^{th}$ eigenvalues. Additionally, $p_{2ijL}$ represents the linear PFs, while $p_{2ijNL}$ is a correction term added to the linear participation [4]. The NPFs is a more accurate measurement of a given state variable's participation in each single mode or mode combination because it considers the second- and higher-order nonlinearities [4], [6].

## III. Proposed Model Reduction Approach

This section introduces the proposed model reduction approach addressing near resonance conditions.

### A. NPFs by Tensor Contraction

Traditional approaches for calculating NPFs often rely on explicit matrix multiplications and involve multiple nested loops for summation. This iterative process significantly increases computational time, making the calculations inefficient for large-scale systems. By leveraging a tensor-based interpretation of NPFs, a highly efficient approach is proposed for computing NPFs, which can be extended to higher-order calculations. Tensor contraction is an operation that reduces the dimensionality of tensors by summing over certain indices. Using Einstein notation, computations are vectorized, making them significantly faster and more efficient by directly contracting tensors without explicitly forming large intermediate matrices. This approach significantly accelerates the calculation of NPFs and other critical stability metrics, making large-scale power system analysis more computationally feasible. The detailed algorithm is given below.

Ref. [30] introduces tensor contraction method to accelerate NPF calculation. However, the approach still relies on loops, which significantly degrade computational efficiency. This paper proposes an improved algorithm by implementing fully vectorized tensor contraction as showed below to obtain $v_{2jii}$ using (5)-(17).

**Fully Vectorized Tensor Contraction Algorithm**

**Require:** $H,\ v,\ u,\ \Lambda,\ n$
**Ensure:** $v_2$
**1. Contract along each dimension:**
**2.** ⋄ $T_1 \leftarrow tensorCon\ (u^T, H)$
**3.** ⋄ $T_2 \leftarrow tensorCon\ (T_1, u)$
**4. Calculate C:**
**5.** ⋄ $C \leftarrow tensorCon\ (T_2, v, 3, 2)/2$
**6. Calculate h2**
**7.** ⋄ ***ones_tensor*** ← *ones(n,n,n)*
**8.** ⋄ $\boldsymbol{\lambda_j} \leftarrow$ ***ones_tensor*** × *reshape(diag(Λ), [n, 1,1])*
**9.** ⋄ $\boldsymbol{\lambda_k} \leftarrow$ ***ones_tensor*** × *reshape(diag(Λ), [1, n,1])*
**10.** ⋄ $\lambda_l \leftarrow$ ***ones_tensor*** × *reshape(diag(Λ), [1, 1,n])*
**11.** ⋄ $D \leftarrow -\boldsymbol{\lambda_j} + \boldsymbol{\lambda_k} + \boldsymbol{\lambda_l}$
**12.** ⋄ $h2 \leftarrow$ C/D
**13. Extract Tensor Elements:**
**14.** ⋄ $p,q \leftarrow$ find $(\text{triu}(1_n))$
**15.** ⋄ $\boldsymbol{v_p} \leftarrow \boldsymbol{v}(\boldsymbol{p},:)$
**16.** ⋄ $\boldsymbol{v_q} \leftarrow \boldsymbol{v}(\boldsymbol{q},:)$
**17. Reshape $\boldsymbol{h2}$ for Contraction:**
**18.** ⋄ $\bar{h}2 \leftarrow -h2(:, sub2ind([n, n], p, q)$
**19. Compute Element-wise Product:**
**20.** ⋄ $\boldsymbol{v_{prod}} \leftarrow v_p . v_q$
**21. Compute Element-wise Product:**
**22.** ⋄ $\boldsymbol{v_2} \leftarrow\ tensorCon\ (\bar{h}2, v_{prod}, 2, 1)$
**23. Return $\boldsymbol{v_2}$**

### *B. Ranking of modes by modal energy*

Modal energy plays a critical role in understanding electro-mechanical and inter-area oscillations, which are characterized
by the periodic interchange of mechanical kinetic energy due to the relative rotor acceleration of generators across the interconnected areas. By analyzing the energy associated with different oscillation modes, valuable insights into the stability of the system can be gained. High modal energy indicates significant oscillatory activity, which can help identify the machines or areas most involved in energy exchange.

To identify the dominant modes, the linearized state-space matrix $A$ is obtained, and the left and right eigenvectors are then calculated. The energy of each mode is determined, and the modes are ranked by their energy levels. This ranking helps identify which modes require further analysis through participation factor calculations. The mode energy for the $j^{th}$ mode is obtained using [31]:

$$E_j = \frac{1}{2} {x_0}^T U_j V_j\, x_0 \tag{19}$$

where $E_j$ is the modal energy associated with the $j^{th}$ system mode and $x_0$ is the initial condition [31]. For any mode, this equation enables the determination of the total modal energy deviation resulting from a specific disturbance or control action. Since the ranking of PFs is based on a single state variable, there is no need to determine the initial values for all state variables. In this study, the initial condition is required only for a key variable, such as generator's frequency for mode ranking.

### *C. Hybrid model reduction: partitioned and unpartitioned strategies*

Traditionally, model reduction is applied to a partitioned system comprising the study area and the external area, as illustrated in Fig. 1. The study area is the primary focus, containing the elements whose detailed behavior is of interest, such as critical components, generators, or converters. The external area is less critical to the specific study but still interacts with the study area. The study area and external area are connected by several tie-lines, and these tie-lines are modeled as fictitious generators. These fictitious generators, acting as constant voltage sources, represent current injections between the areas. During each simulation iteration, the voltage phasors at the boundary buses are recalculated and exchanged as inputs for the next iteration, with each area being calculated separately.

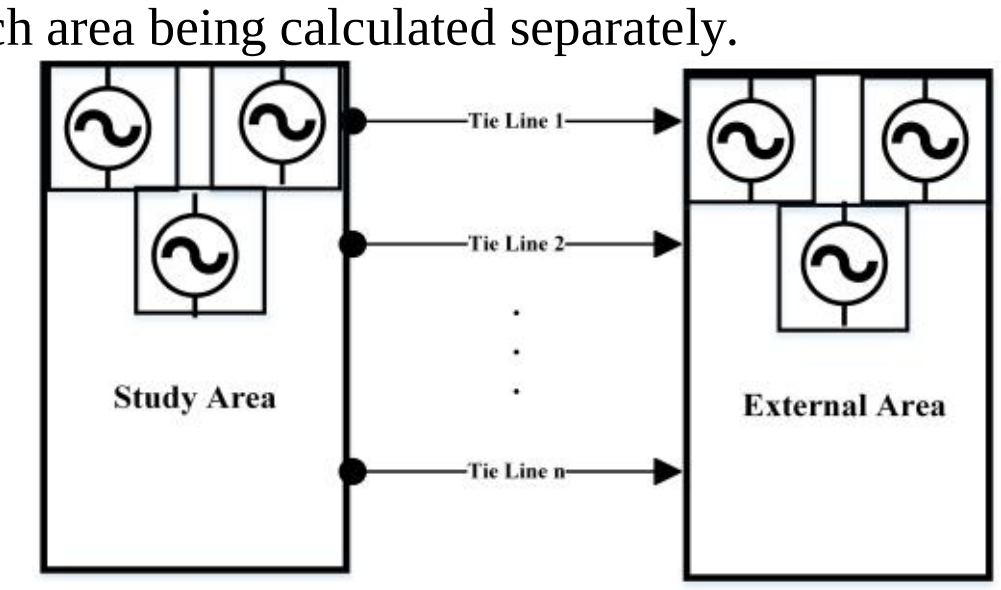

Fig. 1. Partitioned power system model [20]

A model reduction method is proposed as a hybrid of nonlinear and linear model reduction techniques. The balanced truncation method is first used to reduce the system model nonlinearly as:

$$\dot{\tilde{x}} = \begin{cases} \mathbf{T}\ f(\tilde{\mathbf{T}}\tilde{x}, u) \\ y = \tilde{\mathbf{T}}\tilde{x} \end{cases} \tag{20}$$

where T is transformation matrix and $\tilde{T}$ is its inverse. Matrix T transforms the state variables from the original state space to a new balanced state space. In this new balanced system, the state variables are arranged so that the first variable is the most controllable and observable, while the last is the least controllable and observable.

In the proposed hybrid model reduction approach, the functions that contribute least to the dynamics between the external area and the study area are linearized. Therefore, the nonlinear generator functions with small participation factors are linearized, whereas the nonlinear functions corresponding to the generators with large participation factors are kept nonlinear. As a result, the hybrid reduced system can be described as:

$$\dot{\tilde{x}} = \begin{cases} \mathbf{T}\begin{pmatrix} \tilde{f}(\tilde{\mathbf{T}}\tilde{x},\, u) \\ \tilde{\mathbf{A}}\Delta\tilde{x} + \tilde{\mathbf{B}}\Delta u + \hat{x}_0 \end{pmatrix} \\ y = \tilde{\mathbf{T}}\tilde{x} \end{cases} \tag{21}$$

where $\tilde{f}$ includes nonlinear functions and $\tilde{A}\Delta\tilde{x}+\tilde{B}\Delta u+\hat{x}_0$ is the representation of linearized functions. T is transformation matrix and $\tilde{T}$ is its inverse. $\tilde{A}$ and $\tilde{B}$ are:

$$\tilde{A} = \tilde{P}A\tilde{T}, \quad \tilde{B} = \tilde{P}B \tag{22}$$

where $\tilde{A}$ and $\tilde{B}$ are reduced matrices and $\tilde{P}$ is the reduced identity matrix [13].

Dividing the system into the study area and the external area introduces a specific error. This error arises because the inputs, including the boundary bus voltage magnitudes and angles, are computed based on the previous simulation iteration, resulting in a one-iteration lag. To avoid this partitioning error, the proposed adaptive approach can also be implemented on the system without partitioning. Thus, the concept of fictitious generators representing the boundary buses of the study area becomes unnecessary [13].

For the unpartitioned system, the entire system is treated as a single area containing all generators, including those from the study area whose dynamics are of primary interest. Therefore, there is no transformation or truncation of states. The performance enhancement is solely achieved through the linearization of nonlinear functions. Without inputs from the boundary between the study area and the external area, the control matrix $B$ is removed. The representation of this case is as follows:

$$\dot{x} = \begin{cases} \begin{pmatrix} \tilde{f}(x, u) \\ \tilde{\mathbf{A}}\Delta x + \hat{x}_0 \end{pmatrix} \\ y = x \end{cases} \qquad (23)$$

where $\tilde{A}=\tilde{P}A$ and the nonlinear functions representing generators of the study area are included in $\tilde{f}$ [13].

### D. Adaptive Switching Approach

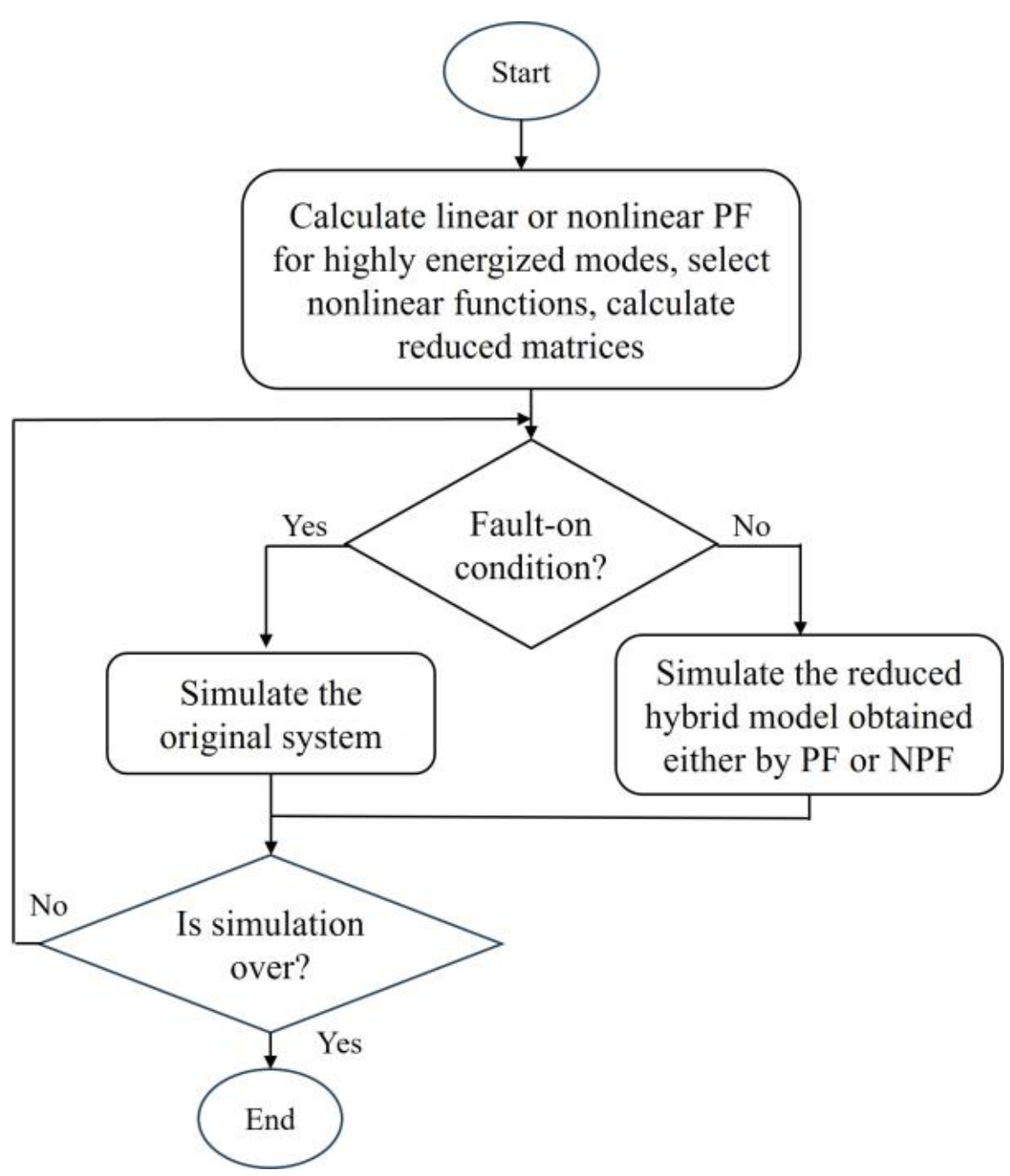

Fig. 2. Flowchart of the proposed approach

A second-order resonance occurs when the sum of two eigenvalues equals a third eigenvalue. The magnitudes of the second-order terms $h_{2jkl}$ in the nonlinear transformation quantify the nonlinear interactions between modes. As shown in (7), the terms $h_{2jkl}$ increase significantly near a second-order resonance, where $\lambda_k + \lambda_l - \lambda_j \simeq 0$. When a resonance or near-resonance condition occurs, the original nonlinear power system, modeled in $x$-space, becomes difficult to transform into a linear system through any nonlinear transformation because of the excessively large $h_{2jkl}$ terms. Therefore, near-resonance condition amplifies the system's nonlinear characteristics around the equilibrium point, making it unsuitable for analysis as a linear system.

In the proposed switching approach, the original fully detailed system model is simulated during the fault-on period to ensure maximum accuracy. Since a fault duration is typically limited to a few cycles, this does not significantly impact the overall simulation time. After the fault, the model switches to a hybrid reduced system, where certain generator functions remain nonlinear while others are linearized based on their PFs. When the system's behavior tends to be nearly linear, the linear participation factor is used to decide which generator models should be kept nonlinear and which ones should be linearized. Conversely, when the system exhibits high nonlinearity or near second order resonance, the nonlinear participation factor is a decision criterion for more accurate selection of generators for linearization. The modal analysis and Normal Form theory presented in equations (1)-(18) are employed to determine the eigenvalues and participation factors of each state variable on the dominant system modes. The flowchart of the proposed method is shown in Fig. 2.

## IV. Case Studies

### A. Test system and environment

An NPCC system shown in Fig. 3 with a 48-machine, 140-bus is used as the test system for case studies.

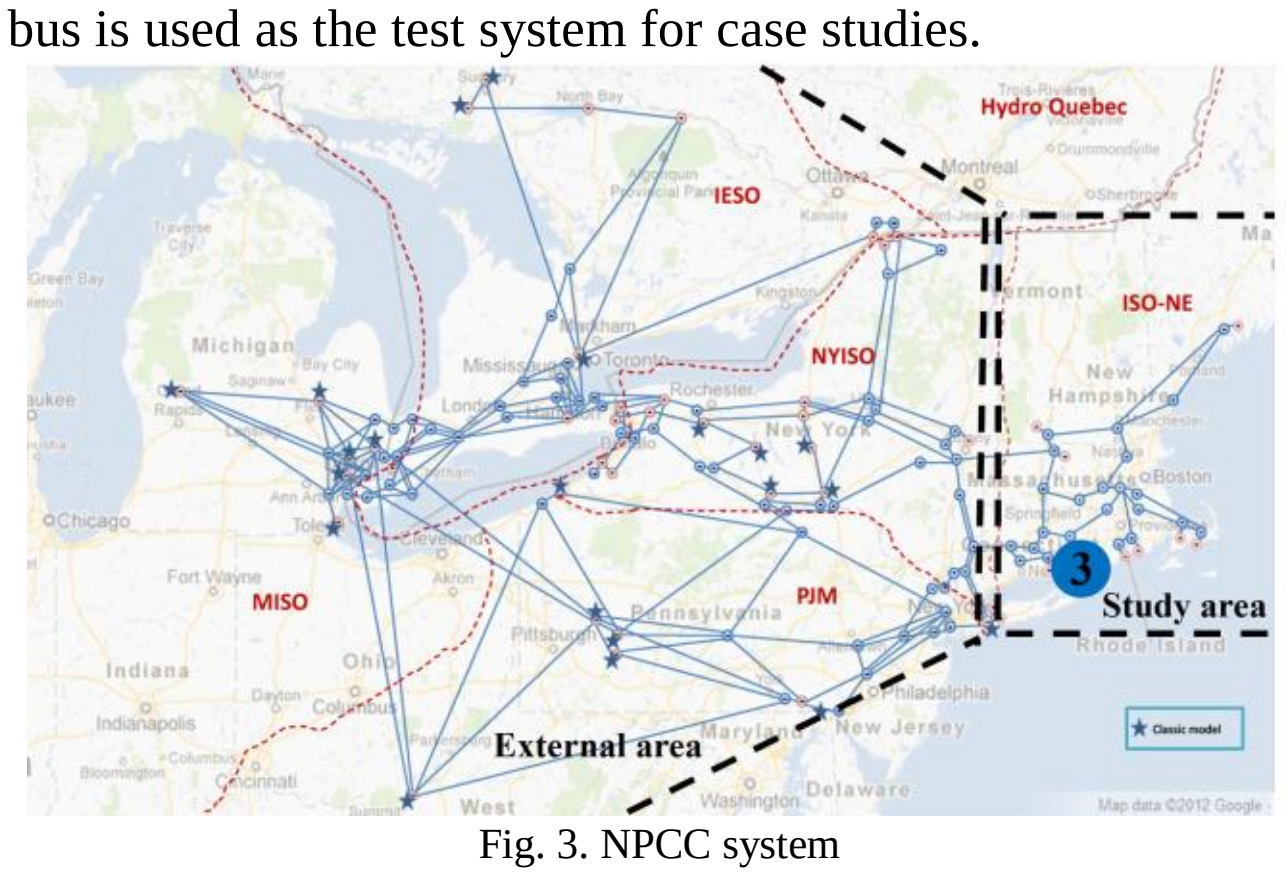

Fig. 3. NPCC system

In this system, each generator and associated controllers are modeled by nine first-order differential equations including a detailed two-axis generator model, a first-order governor model, a non-reheat steam turbine model, and an IEEE type-1 exciter. The state vector and input vector are:

$$x = \left(\delta\ P_m\ P_{gv}\ V_R\ R_f\ E_{fd}\ E'_d\ E'_q\ \omega\right) \qquad (24)$$

$$u = \left(\theta\ V\right) \qquad (25)$$

where $\delta$ and $\omega$ denote the rotor angle in rad and the speed of generators in rad/s, respectively. $P_m$ is the mechanical power, $P_{gv}$ is the governor output power, $V_R$ shows voltage regulator input and $R_f$ is rate feedback. $E_{fd}$, $\acute{E}_q$, $\acute{E}_d$ are field voltage, internal voltages on the q-axis and the d-axis, respectively. $\theta$

and $V$ are the voltage angle and voltage magnitude at boundary buses [13], [20]. The fault is at bus 3 as shown on the map in Fig. 3. Generators 27 at bus 78 is the reference generator and generator 5 at bus 23 is the output generator.

All simulations are carried in MATLAB 2024b on a PC with Intel(R) Core(TM) i7-10700 CPU@ 2.90GHz 16 GB processor. The model reduction is done for partitioned and unpartitioned systems under near resonance conditions as explained in the following sections.

### *B. Partitioned strategy*

The study area consists of nine generators in the ISO-NE region, comprising 81 state variables, while the external area includes 39 generators with 351 state variables. These two regions are connected through two tie-lines. The study area retains its original, detailed models, whereas the external area is designated for model reduction.

The NPCC system is analyzed using modal analysis to identify near-resonance scenarios and the associated modes. Table I highlights two resonance cases. In the first case, resonance occurs due to modes with indices 69, 58, and 21, with frequencies of 0.7184 Hz, 0.948 Hz and 1.663 Hz, while in the second case, resonance is formed by modes with indices 69, 64, and 31 with frequencies of 0.7184 Hz, 0.8132 Hz and 1.550 Hz. Mode 69 with a frequency of 0.7184 Hz, appears in both instances, indicating its significant nonlinear interaction.

TABLE I
RESONANCE CASES-PARTITIONED SYSTEM

| | 1th Resonance | | 2th Resonance | |
|---|---|---|---|---|
| | modes | index | modes | index |
| $\lambda_k$ | -0.3239 + 4.5139i | 69 | -0.3239 + 4.5139i | 69 |
| $\lambda_l$ | -0.3431 + 5.9567i | 58 | -0.3868 + 5.1099i | 64 |
| $\lambda_j$ | -0.7097 + 10.4504i | 21 | -0.7272 + 9.7398i | 31 |
| $\lambda_k + \lambda_l - \lambda_j$ | 0.0426 + 0.0202i | | 0.0165 + 0.1158i | |

The energies of the modes are calculated using (19) and included in Fig. 4. As can be seen, the energy of mode number 69 involved in the resonance, is the highest and therefore the linear and nonlinear PFs are calculated for this mode.

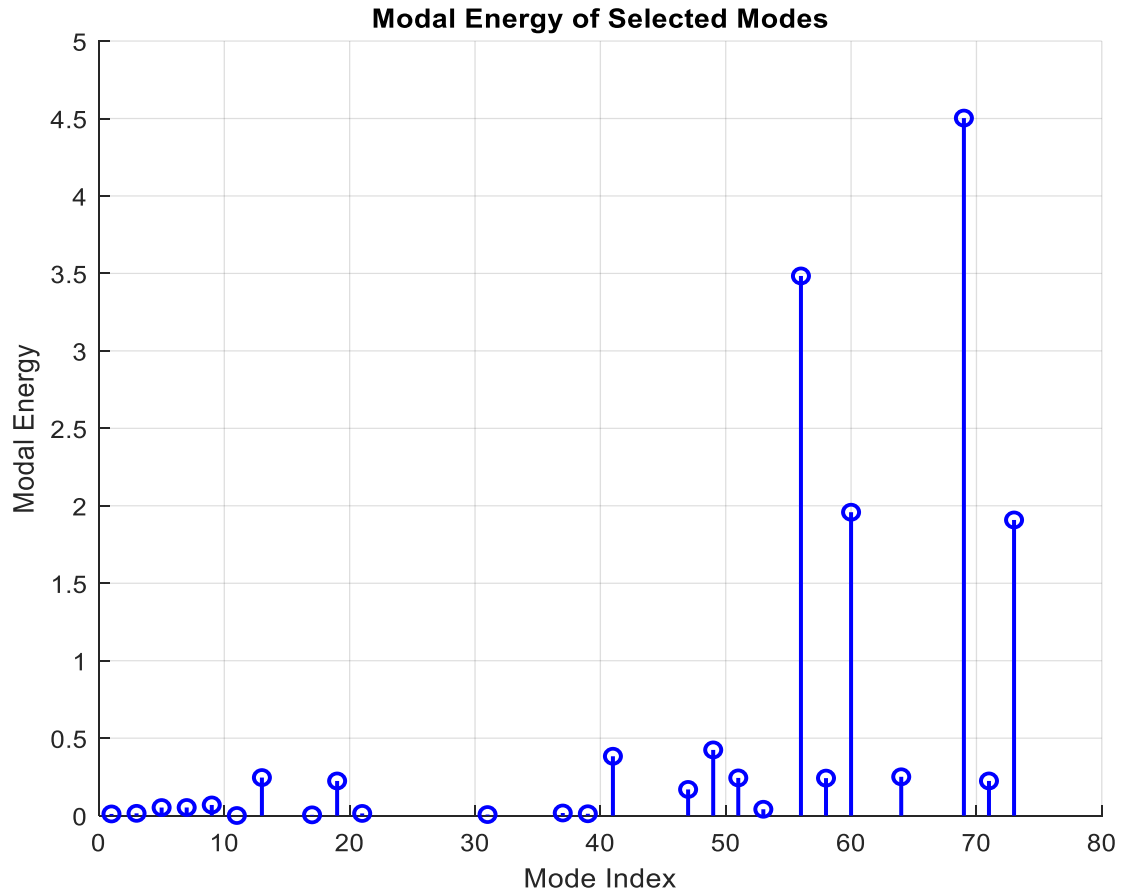

Fig. 4. Energy of modes-partitioned system

Calculating the second-order numerical Hessian matrix through perturbations around the equilibrium points takes 11.96 seconds, significantly faster than the analytical approach, which requires several hours. Table II compares the time needed to calculate nonlinear PFs in both partitioned and unpartitioned system using traditional loop-based approach, tensor method in [30] and the proposed tensor approach in algorithm 1. The proposed tensor method significantly speeds up the NPF calculation compared to the previous methods.

TABLE II
NPF CALCULATION TIME

| NPF calculation time | Partitioned | Unpartitioned |
|---|---|---|
| Full loop (traditional method) | > 24 hours | > 24 hours |
| Tensor contraction (old) | 109.811 sec | 266.454 sec |
| Tensor contraction (accelerated) | 3.341 sec | 12.031 sec |

By applying the same threshold to the linear and nonlinear PFs, the generators with the highest PF values are identified and modeled nonlinearly, while the remaining generators are linearized. In this case, 12 generators are chosen to remain nonlinear based on nonlinear PF case and 13 generators based on the linear PF case. These selected generators with higher participation than the threshold are illustrated in Fig. 5.

From the PF results in Fig. 5, the generators with the highest participation in the resonance mode differ between the nonlinear and linear PFs. Generators 27, 31, 41, and 44 exhibit significant participation according to the nonlinear PFs but show negligible participation in the linear PFs. In contrast, generators 13, 15, 19, 20, and 22 are selected based on the linear PFs. The remaining selected generators are the same.

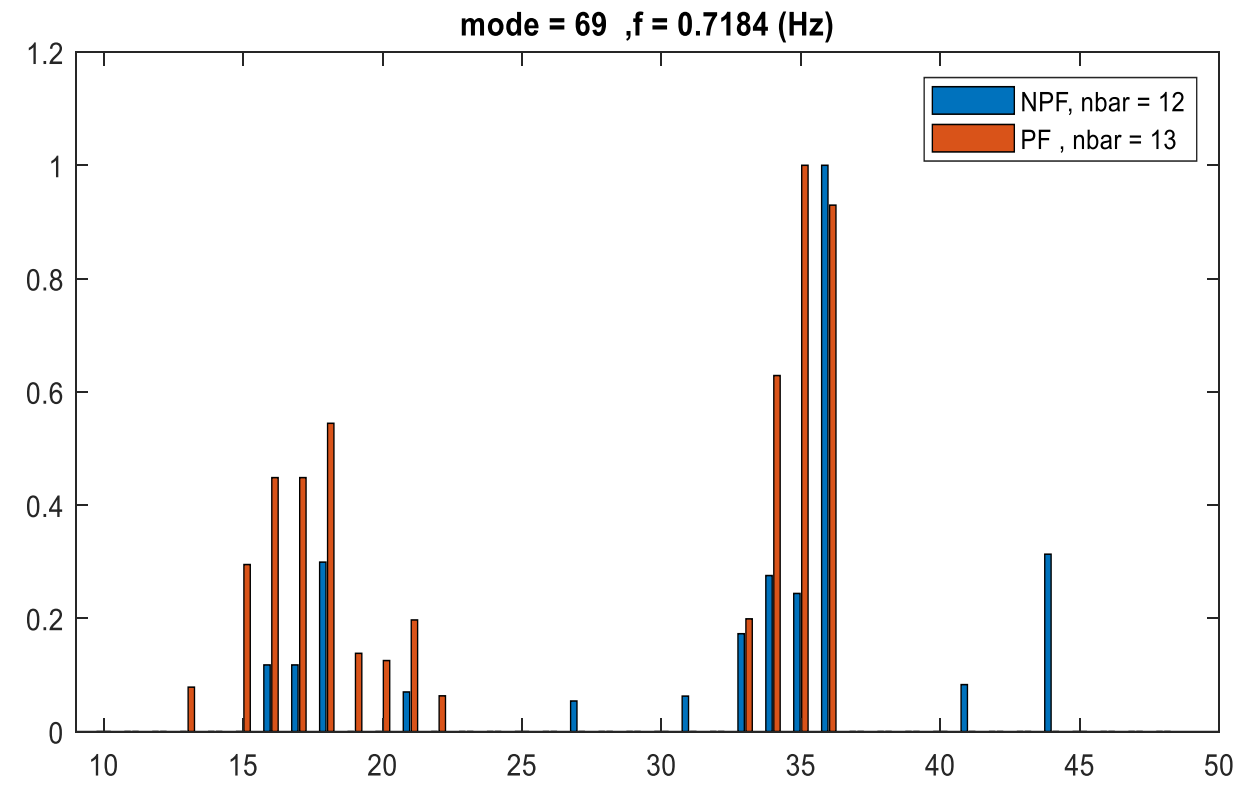

Fig. 5. Nonlinear and linear PFs for mode 69

The fault at bus 3 is the largest disturbance, with generator 5 exhibiting the highest rotor angle error. Thus, generator 5's rotor angle is used for comparison. In this study, a three-phase short circuit fault with a duration of 0.39 seconds is applied to bus 3 of the NPCC system, and the overall simulation time is 16 seconds. The fault duration is the critical clearing time (CCT) required to maintain the system stability. A fault with CCT represents the worst-case scenario, leading to the largest error. The simulation time step is considered 0.01 seconds. Fig. 6 shows the rotor angle responses for generator 5 for the original system, a hybrid model based on PF, and a hybrid model based on NPF, and its reduced version using balanced truncation. As seen in the figure, the reduced model by nonlinear PFs closely follows the original system response and provides less error compared to the reduced model obtained by linear PFs.

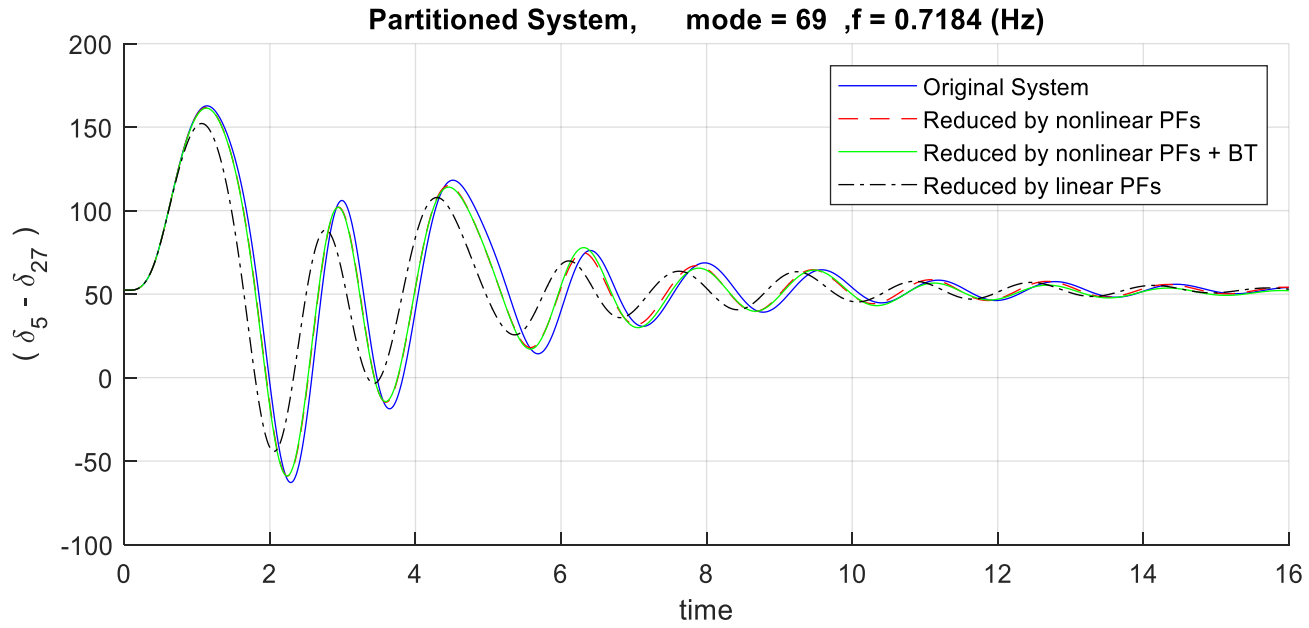

Fig. 6. Rotor angle of generator 5 following a fault at bus 3

The root mean squared error (RMSE) of state variables are calculated for generator 5 using (26) and they are included in Table III.

$$\varepsilon_i = \sqrt{\frac{\sum_{j=1}^{n}(x_{ij} - \hat{x}_{ij})^2}{n}} \tag{26}$$

where $n$ is the number of time steps, $x_{ij}$ is the $i^{th}$ state variable at time step $j$ and $\hat{x}_{ij}$ indicates the $i^{th}$ state variable for the reduced-order model at time step $j$.

TABLE III
RMSE OF STATE VARIABLES FOR GENERATOR 5

| State variables | Approaches | | |
|---|---|---|---|
| | Linear PFs | NPFs | NPFs+ BT |
| $\delta$, degrees | $22.90\times10^{0}$ | $6.277\times10^{0}$ | $6.501\times10^{0}$ |
| $P_m$, p.u. | $1.978\times10^{-3}$ | $1.218\times10^{-3}$ | $3.531\times10^{-3}$ |
| $P_{gv}$, p.u. | $1.705\times10^{-2}$ | $5.083\times10^{-3}$ | $8.046\times10^{-3}$ |
| $V_R$, p.u. | $1.471\times10^{-1}$ | $4.190\times10^{-2}$ | $4.458\times10^{-2}$ |
| $R_f$, p.u. | $1.113\times10^{-2}$ | $3.107\times10^{-3}$ | 3.306×10-3 |
| $E_{fd}$, p.u. | $8.519\times10^{-2}$ | $2.421\times10^{-2}$ | $2.560\times10^{-2}$ |
| $\acute{E}_d$, p.u. | $6.151\times10^{-2}$ | $1.675\times10^{-2}$ | $1.829\times10^{-2}$ |
| $\acute{E}_q$, p.u. | $9.486\times10^{-3}$ | $2.661\times10^{-3}$ | $2.833\times10^{-3}$ |
| $\omega$, p.u. | $3.642\times10^{-3}$ | $1.012\times10^{-3}$ | $1.166\times10^{-3}$ |

TABLE IV
SIMULATION TIME COST

| System | Simulation time (Sec) |
|---|---|
| Original full order | 0.6011 |
| Partitioned linear PF-based | 0.2586 |
| Partitioned nonlinear PF-based | 0.2473 |
| Partitioned reduced nonlinear PF-based | 0.2174 |

The rotor angle state variable consistently exhibited the largest error for each generator. As Table III shows the reduced model by linear PFs resulting in errors exceeding 20 degrees. In contrast, the proposed NPFs approach limits the error to 7 degrees. The simulation time cost for the original and the reduced models is included in Table IV.

Given that PFs are independent of both the specific initial conditions and the nature of disturbances within the system, linearization and PFs calculation are performed offline in this study, with updates required only when there is a significant change in the operating conditions. As a result, the need to recalculate matrices does not impact on the algorithm's speed over extended periods.

## C. Unpartitioned strategy

This section explores the application of the proposed approach to an unpartitioned NPCC system. In the unpartitioned case, the system is treated as a whole, comprising 432 state variables. As shown in Table V, two resonance cases are identified in the unpartitioned case, both involving mode 88 with a frequency of 0.7193 Hz. Fig. 7 shows the modal energy for different modes, from which mode 88 involved in the resonance shows the largest energy.

The numerical Hessian matrix takes 26.01 sec to be calculated and the NPF calculation time is included in Table II. Applying a threshold, identifies 13 key generators with the highest PF values in both NPF and linear PF approaches to remain nonlinear, while the rest are linearized. The nonlinear and linear PFs higher than the threshold for mode 88 are presented in Figure 8.

Different generators are selected for nonlinear modeling in the linear PF and nonlinear PF approaches, showing the varying influence of mode 88 across the two methods.

TABLE V
RESONANCE CASES-UNPARTITIONED SYSTEM

| | 1th Resonance | | 2th Resonance | |
|---|---|---|---|---|
| | modes | index | modes | index |
| $\lambda_k$ | -0.3260 + 4.5192i | 88 | -0.3195 + 3.9613i | 92 |
| $\lambda_l$ | -0.3431 + 5.9562i | 76 | -0.3260 + 4.5192i | 88 |
| $\lambda_j$ | -0.7098 + 10.4504i | 25 | -0.6527 + 8.5605i | 49 |
| $\lambda_k+\lambda_l-\lambda_j$ | 0.0406 + 0.0251i | | 0.0071 + 0.0799i | |

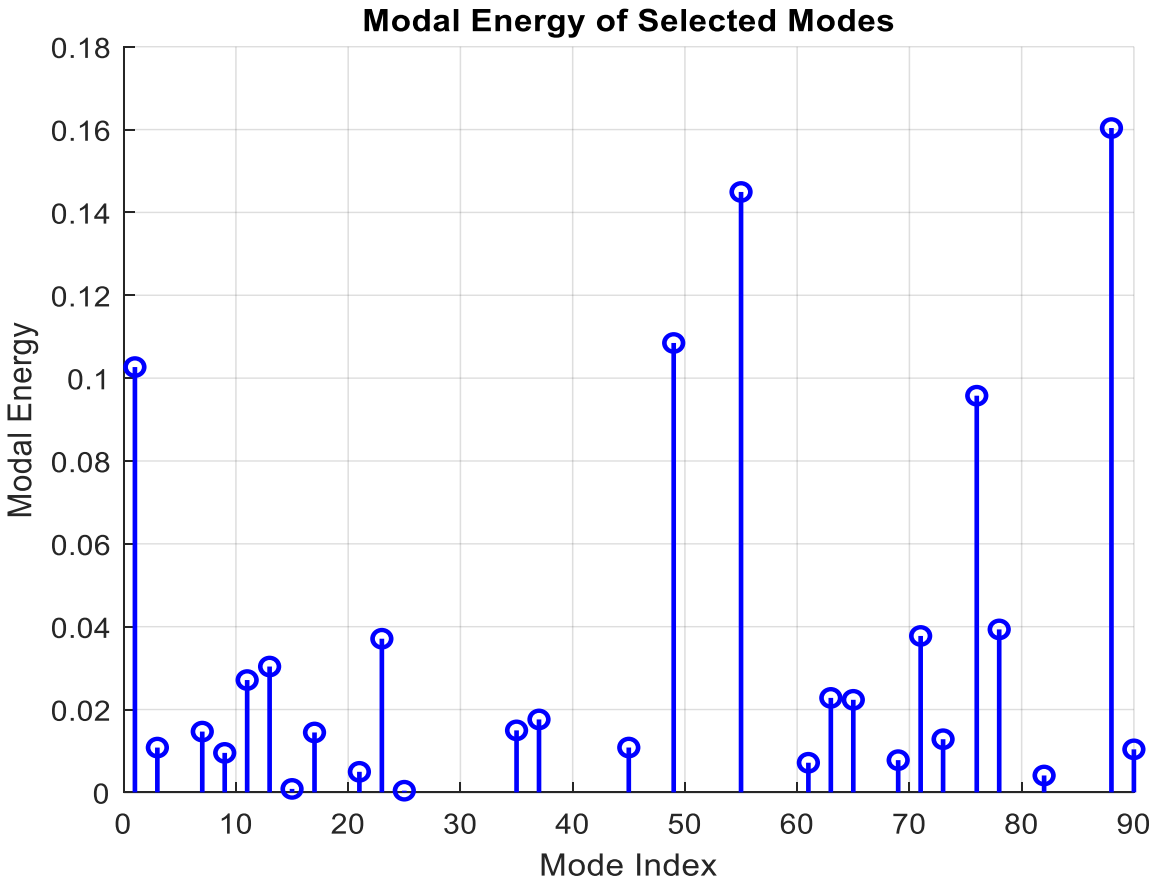

Fig. 7. Modal Energy-Unpartitioned case

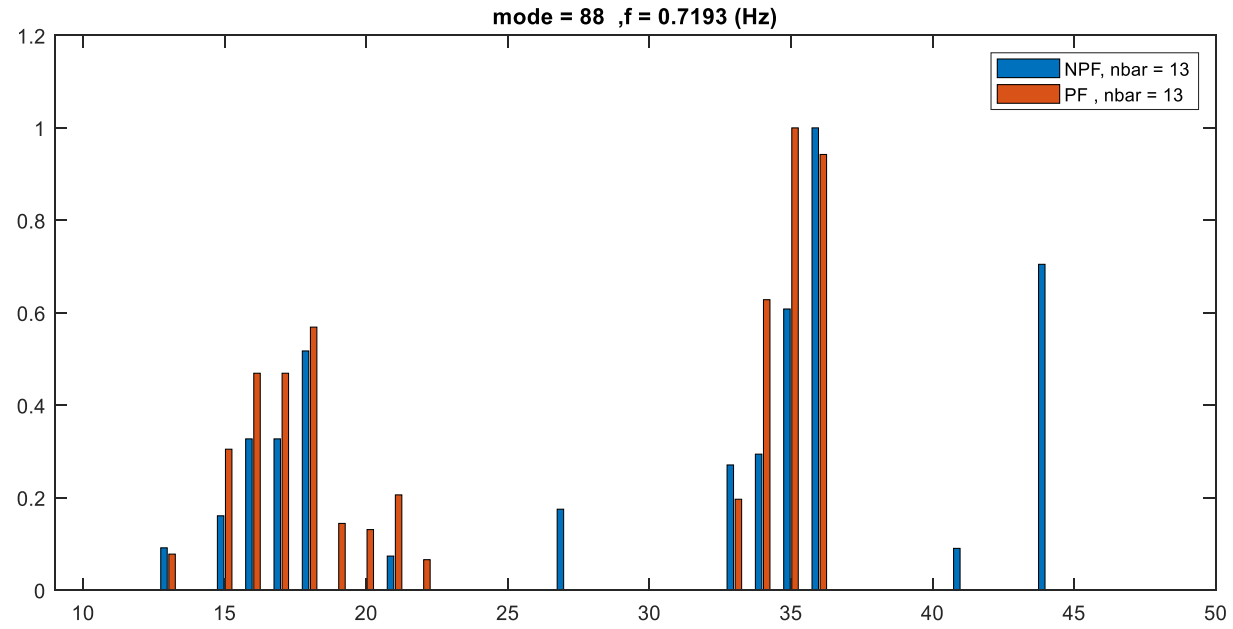

Fig. 8. Nonlinear and linear PFs for mode 88

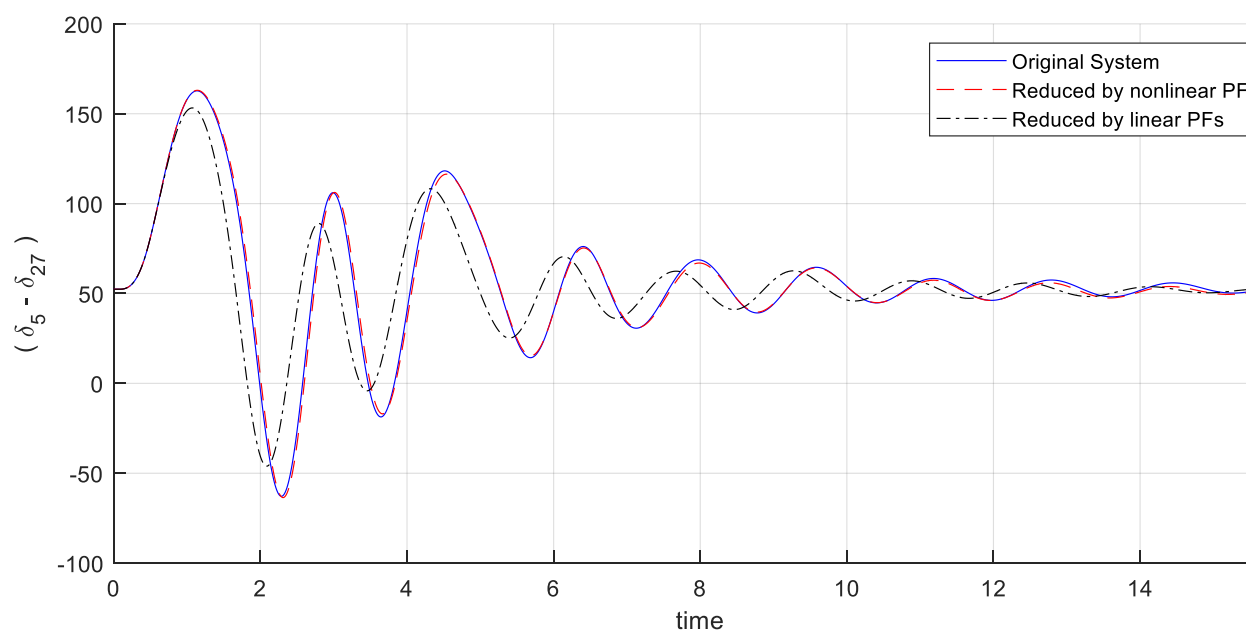

Fig. 9. Rotor angle of generator 5 following a fault at bus 3

Fig. 9 illustrates the rotor angle responses for generator 5 in both the original system and the reduced models using nonlinear and linear participation factors. The figure shows that the reduced model based on nonlinear participation factors closely matches the original system's response, resulting in smaller errors compared to the model that reduced using linear participation factors.

TABLE VI
RMSE OF STATE VARIABLES FOR GENERATOR 5

| State variables | Approaches | |
|---|---|---|
| | ***Linear PFs*** | ***NPFs*** |
| $\delta$, degrees | $20.41\times10^{0}$ | $2.442\times10^{0}$ |
| $P_m$, p.u. | $1.534\times10^{-3}$ | $1.142\times10^{-3}$ |
| $P_{gv}$, p.u. | $1.510\times10^{-2}$ | $2.753\times10^{-3}$ |
| $V_R$, p.u. | $1.308\times10^{-1}$ | $2.842\times10^{-2}$ |
| $R_f$, p.u. | $9.784\times10^{-3}$ | $2.210\times10^{-3}$ |
| $E_{fd}$, p.u. | $7.509\times10^{-2}$ | $1.643\times10^{-2}$ |
| $\acute{E}_d$, p.u. | $5.498\times10^{-2}$ | $6.759\times10^{-3}$ |
| $\acute{E}_q$, p.u. | $8.544\times10^{-3}$ | $2.695\times10^{-3}$ |
| $\omega$, p.u. | $3.240\times10^{-3}$ | $4.224\times10^{-4}$ |

The RMSE of the state variables for generator 5 is calculated and presented in Table VI, demonstrating that the approach using nonlinear participation factors has significantly reduced the error for state variables compared to the linear participation factor approach. Specifically, the rotor angle error is 2.44° for the hybrid model based on NPF, whereas it is 20.41° for the model based on PF. Therefore, NPF offers a better selection of key generators under resonance conditions. The simulation time cost for both the original and reduced models is provided in Table VII, showing that the simulation time for the nonlinear participation factor approach is in the same range as the linear approach and is reduced compared to the full nonlinear model. The results indicate that under high nonlinearity conditions, linear PFs analysis might not provide an accurate modal characteristic of the power system.

TABLE VII
SIMULATION TIME COST

| System | Simulation time (Sec) |
|---|---|
| Original full order | 0.6011 |
| Unpartitioned linear PF-based | 0.3134 |
| Unpartitioned nonlinear PF-based | 0.3279 |

## V. CONCLUSION

To address high nonlinearity conditions, such as those occurring near resonance, this paper presented a new hybrid adaptive approach to accelerate power system simulation using nonlinear PFs. Tensor contraction and the numerical Hessian matrix method are employed to significantly accelerate the computation of nonlinear PFs compared to conventional approaches. The method is implemented on the NPCC 140-bus system under two cases: partitioned and unpartitioned system. Under the near resonance case, the second order nonlinearity coefficients become very large, resulting in a considerable difference between linear and nonlinear PFs. As such, the reduced model obtained by nonlinear PFs provides better representation of the original system compared to the linear PFs approach.